\documentclass[doublespacing]{bmcart}
\usepackage{graphicx}
\usepackage{multirow}

\usepackage{algpseudocode}
\usepackage{hyperref}
\usepackage{color,soul}
\usepackage{graphicx}
\usepackage{lscape}
\usepackage{adjustbox}
\usepackage{tabularx}
\usepackage[T1]{fontenc}
\usepackage{hyperref}
\usepackage{cite}
\usepackage{amsmath,amssymb,amsfonts}
\usepackage{algorithm}
\usepackage{textcomp}
\usepackage{xcolor}
\usepackage{pgfplots}
\usepackage{pgfplotstable}
\pgfplotsset{compat=1.7}
\usepackage{csvsimple}
\usepackage{rotating}
\usepackage[cmintegrals]{newtxmath}
\usepackage{longtable}
\usepackage{colortbl}
\usepackage{babel}
\usepackage{authblk}
\usepackage{blindtext}
\usepackage{float}
\usepackage{dirtytalk}
\usepackage[document]{ragged2e}
\usepackage[justification=centering]{caption}
\usepackage{mathtools,amssymb,lipsum}

\usepackage{cuted}
\usepackage{wrapfig}
\RequirePackage{natbib}

\begin{document}

\begin{frontmatter}

\begin{fmbox}
\dochead{Research}


\title{Generative Adversarial Networks for Malware Detection: a Survey}


\author[
    email={a.dunmore@massey.ac.nz}
]{Aeryn Dunmore$^1$}
\author[
    email={j.jang-jacquard@massey.ac.nz}
]{\inits{JJJ}\fnm{Julian} \snm{Jang-Jaccard$^1$}}
\author[
    email={f.sabrina@cqu.edu.au}
]{\inits{FS}\fnm{Fariza} \snm{Sabrina$^2$}}
\author[
    email={security@ajou.ac.kr}
]{\inits{JK}\fnm{Jin} \snm{Kwak$^3$}}
\address[id=aff1]{%
  \orgname{Cybersecurity Lab, Massey University}, 
  \city{Auckland},                              
  \cny{NZ}                                    
}
\address[id=aff2]{%
  \orgname{School of Engineering and Technology, Central Queensland University},
  \city{Sydney, NSW},
  \cny{Australia}
}
\address[id=aff3]{%
  \orgname{Department of Cyber Security, Ajou University},
  \city{Suwon},
  \cny{Republic of Korea}
}
\begin{artnotes}
\small
    \\$^1$Cybersecurity Lab, Massey University, Albany, New Zealand\\
    $^2$School of Engineering and Technology, Central Queensland University, Sydney, NSW, Australia\\
    $^3$Department of Cyber Security, Ajou University, Suwon, Republic of Korea\\
\end{artnotes}

\end{fmbox}


\begin{abstractbox}

\begin{abstract} 
Since their proposal in the 2014 paper by Ian Goodfellow \cite{goodfellow2014generative}, there has been an explosion of research into the area of Generative Adversarial Networks. While they have been utilised in many fields, the realm of malware research is a problem space in which GANs have taken root. From balancing datasets to creating unseen examples in rare classes, GAN models offer extensive opportunities for application. This paper surveys the current research and literature for the use of Generative Adversarial Networks in the malware problem space. This is done with the hope that the reader may be able to gain an overall understanding as to what the Generative Adversarial model provides for this field, and for what areas within malware research it is best utilised. It covers the current related surveys, the different categories of GAN, and gives the outcomes of recent research into optimising GANs for different topics, as well as future directions for exploration.\\
\end{abstract}




\end{abstractbox}
%

\end{frontmatter}


\renewcommand{\thefootnote}{\roman{footnote}}
\justifying
\section{Introduction}\label{introduction}
Generative Adversarial Networks, or GANs, are a type of deep learning neural network model based on the Game Theory premise of a zero-sum game\cite{goodfellow2014generative}. These networks have become popular in many fields as a Machine Learning (ML) model which has great success at synthesising large samples of dataset classes based on learning classes and features from an existing dataset. They are particularly good at synthesising images \cite{Jang20201}, making them both popular in computer vision tasks, and excellent at generating malware 'images' for training systems to detect malicious files and applications. They also offer a chance to augment the rarest classes in a dataset \cite{Pham.2021}. While they have received attention from many disciplines and research topics, the research into GANs for synthesising images of different malware classes is very promising, and as such, is where we have chosen to focus our survey. 
\newline
The research in this study concerns of the state of the art in GANs and Malware, and we have found a space for an up-to-date examination of where this research discipline is and where it appears to be headed. As is explained in Section \ref{relatedwork}, on related works, while there are surveys or studies which are \textit{similar} in aspects of their research, to our knowledge there is no updated survey on this topic. This is of distinct importance because of how much growth we have seen in this area of research in recent years. We have done our utmost to present a balanced examination, with both breadth and depth, which can be of use to both researchers new to the area, and those wanting an update on the current problem space. We have also attempted to approach this survey in a way that makes it accessible for the machine learning or cybersecurity researcher both. 
\newline
The rest of the paper is structured as follows: Section \ref{whatisgan} describes the structure of GAN models, and how they are built and trained, as well as defining malware for the purposes of this paper. Section \ref{relatedwork} gives a relatively brief breakdown on the recent works with the most similarities to our survey, and explains how we have developed something different in content and structure. Section \ref{metrics} clarifies the different methods by which researchers measure the performance of their respective GAN models. Section \ref{dataset} explains the datasets used in the experiments and research we have surveyed and what types of data they contain as well as their origins. Section \ref{GANs} delves into the different types of Generative Adversarial Network - both the most commonly implemented models and the innovations that have come from recent researchers developing new ways in which to use the model. Section \ref{areasofuse} will explain the types of uses GAN models are being used in malware research, along with the specific areas within the area to which GAN research is contributing. Section \ref{discussion} contains our in-depth discussion of how GANs are functioning within malware research and what this means for researchers both in cybersecurity and machine learning. Finally, Section \ref{futureresearch} discusses the opportunities for future research, and then Section \ref{conclusion} goes on to the conclusions we believe can be drawn from the survey of papers within this discipline.

\section{Terms, Definitions, and Explanations} \label{whatisgan}
A Generative Adversarial Network, at its base, is a machine learning algorithm built out of two separate deep learning networks which work together, competing to win a zero-sum game. One network takes in noise and then attempts to create samples with the right characteristics to have them seem like real or 'genuine' samples. The second network takes as input both real and generated samples, and then classifies them as either real or fake samples. The back-propagation that occurs then is the backbone of the model. If the Discriminator network is right, information is sent back to the Generator network, so that it will adjust its weights and probability distributions to improve the quality of its forgeries. If the Discriminator instead gets it wrong, that information is sent to the Discriminator to make the necessary adjustments. The games end when the Discriminator has the accuracy of a coin flip - when the forgeries are all but impossible to separate from the genuine samples.
\newline
As discussed above, the Generator creates data that is meant to look as real as possible. The Discriminator has only one job - determine if the data provided to it is generated data (created by the Generator) or if it is in fact genuine data. The Generator is considered to have "won" when the Discriminator has a success rate of 50\%. Meaning that the Generator is so good at producing almost real data, that the Discriminator is left with the same accuracy as tossing a coin. The GAN is considered an unsupervised machine learning method, and it was developed in 2013 to help model the behaviour of wildlife\cite{Aggarwal.2021}. GANs are an alternative generative model to Variational Autoencoders (VAE), which can also be used to create new samples from a given dataset. In many of the papers we surveyed, VAE were used as a point of comparison/control in the experiments for improved GAN models.
\subsection{How does a GAN model work?}\label{howganworks}
Having given a simplistic overview, we now explain the architecture of the GAN model for machine learning in detail. The architecture is innovative for the way it processes and creates both information and datasets. In a world where we need exceptionally large datasets to train the machine learning algorithms that are now slipping into so much of our technology - and therefore into our lives - the ability to \textit{create} new data for training purposes is invaluable. This is, of course, provided the data is, or is at least almost, authentic. Creating exceptional forgeries, in the way that GANs do, is therefore a reason in and of itself to employ them in most problem domains. Cybersecurity especially has need of as many samples as possible of the different types of malware, in order to be able to train defensive technologies, to detect when a file or action is malicious. 
\begin{figure*}
    \centering
    \includegraphics[width=\textwidth]{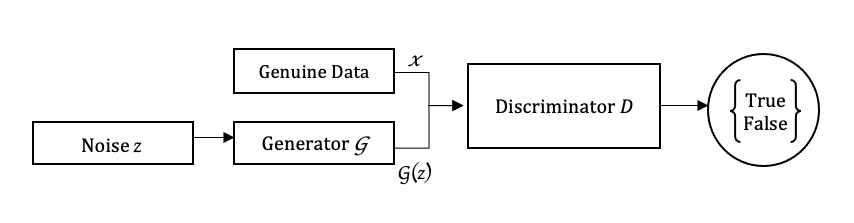}
    \caption{A standard GAN, SGAN, or LSGAN model, reproduced based on diagrams from Wang et al., 2017 \cite{wang2017generative}}
    \label{fig:gan_structure}
\end{figure*}
\newline

\subsection{A Standard GAN Model Structure}\label{implemtation}
The standard/vanilla Goodfellow GAN design is simple but ingenious. It relies on the use of two main deep learning neural nets, with back-propagation and feedback. The Generator and Discriminator play a "two-player minimax game" \cite{goodfellow2014generative} using the value function found in Equation \ref{optimal discriminator}. This equation is taken from Goodfellow's original 2014 paper\cite{goodfellow2014generative} introducing Generative Adversarial Networks\cite{goodfellow2014generative}.\newline
Figure \ref{fig:gan_structure}, from \cite{wang2017generative}, shows the structure of a standard model GAN, and one of its derivatives, Least Squares GAN (LSGAN) discussed later in Section \ref{lsgan}. \newline

\subsubsection{Generator}\label{Generator}
The Generator portion of the network is arguably the more complex. The Generator Network within the GAN model starts with a random seed or noise as input, and produces an output which starts understandably far from the goal. However, as the Discriminator Network feeds information through back-propagation to the Generator, it slowly achieves convergence on the target samples. Each iteration, each epoch, the Generator syntheses data that is more and more realistic. Convergence and training of the Generator Network is finished when the Discriminator cannot tell if the synthesised data is real or manufactured with any more than 50\% accuracy - effectively becoming as useful in classification as a coin toss.
The advantages to being able to do this are many, and this generation of data is what makes the GAN model so popular in the world of machine learning. The ability to extract and analyse important features of each dataset class are a highly sought after trait in machine learning research. 
\newline

\subsubsection{Discriminator}\label{Discriminator}
The Discriminator's job essentially comes down to a binary option. Is the data real, or is it synthesised? This part of the network, like most deep neural networks, is about feedback - in this case, the back-propagation of the result to the Generator Network. When the Discriminator guesses incorrectly, that alters the internal weights of the system as the information back-propagates to both Generator and Discriminator. The Discriminator is also playing to lose, as the target is to have the Discriminator as effective at distinguishing real data from generated data as a simple coin toss. At that point in the game, the Generator has graduated to building data for use in other scenarios. The data provided is considered so close in feature-space that it may as well be treated like the real deal. 
The Discriminator is the part of the dual network that recognises when the model has been sufficiently trained to produce high-quality synthesised data. It is, also, no longer required once the Generator is producing the samples at that level of 'perfection'. It does not need to be used outside of the training of the Generator, as that is its singular purpose.

\subsubsection{Types of feedback loops}\label{typesoffeedbackloops}
Essential to every GAN implementation is the type of feedback response network it utilises. Being able to feed the results of the tests run by the Discriminator back to the Generator is key in the development of an effective Generator. With every wrong choice, the Generator gets to adjust the weights that little bit closer to where they want to be. Likewise, the Discriminator choosing correctly means that the Generator needs to update its neural weights differently, to change the emphasis on particular neurons, and to change the outputs of the network for a more favourable outcome. The back-propagation of the network allows both systems to carefully adjust the weights and probabilities of their internal deep learning neural networks. Because of the feedback loop needed by GAN architecture, it is a back-propagation network. There are networks that change the types of feedback loops and what is presented to cause changes in the algorithm. The function to alter the weights is essential to the training period, and is as such customised in different models and applications such that the network can reach optimal function as efficiently as possible. It is important to remember that this adjustment over time occurs in a black-box. The results and inputs are what we as researchers can control, while the actual operations can only be modelled and estimated.

\subsection{Malware}
Malware, in the general definition for the purposes of this paper, is code written to cause malicious behaviour. Trojans and ransomware are both examples of malware. In its purest form, malware is programming intended to in some way break or disrupt the regular operation of an operating system \cite{Amin.2022}. One of the most well-known malware attacks is WannaCry's ransomware in 2017, infecting computers across the globe with hackers locking users out of their PCs. As ransomware has evolved, it has become a billion dollar business \cite{O'Kane.2018}. This is simply another reason to use every means at our disposal to create a new generation of detection systems, such that individuals (who may be mostly unaware of how best to protect themselves) need not attempt to protect their devices on their own. This paper aims to show the ways in which GAN models can and are being used in order to help create these new systems.

\section{Related Work}\label{relatedwork}

The areas in which the use of GAN is both possible and implemented is exhaustive. The classification, generalisation, and feature extraction abilities of the GAN models make them useful in too many fields to reasonably keep track of, but there are many surveys that have tried to enumerate all the ways in which GAN models helped in their fields. These include steganography \cite{Tang.2017}, the cracking of cryptographic methods \cite{Corley.2019}, e-commerce \cite{Kumar.2018}, and cross-lingual methods for detecting hackers \cite{7y}. There are however, some areas in which GAN models are more common than others. We have summarised the types of GAN models examined in the related works in Table \ref{table:relatedresearch}.
\newline

In Berman et al., 2019\cite{Berman2019}, the authors take the chance to explore the ways in which deep learning methods have been integrated into different realms of cybersecurity. Our paper is similar, though it focuses specifically on the work done using Generative Adversarial Networks for implementation in the research space of malware, both generation and detection. Berman et al. \cite{Berman2019} is intended to familiarise the reader with research into machine learning for cybersecurity. The authors are careful to emphasize the difference between the deep versus shallow neural network machine learning models. This survey is far greater in depth than spread - the types of uses for GAN models in the survey are a small number primarily about identifying attacks, but they are all examined in great detail over the course of many papers in each area. Their survey is indeed intended to familiarise the reader first with machine learning, neural networks, and deep learning, before moving on to Generative Adversarial Networks and then to their applications in cybersecurity. In this, the paper does achieve its aims. However, the lack of variety in the tasks in which GAN models can be used in cybersecurity is a note on which there is certainly room for more breadth. This survey is very careful to ensure the reader understands the metrics, and the results for each of the different papers included in the survey are clearly outlined. The metrics are important and effective at communicating why researchers should further investigate GAN methods for different tasks. 
\newline

In Liu, et al., 2022, \cite{Liu2022123}, the authors undertake a thorough, extensive, and careful examination of the state of the art in Adversarial Machine Learning - of which GANs are one type. The stated purpose of this survey is to examine the weaknesses of Machine Learning Intrusion Detection Systems, and look at ways AML models are being implemented to assist in defending them. The over-arching theme, of course, is through the creation of adversarial examples such that Machine Learning approaches can train on unseen data, and learn to spot the necessary contextual and semantic relationships that can signal malicious code.\newline 
Like our paper, it includes a table of related and similar works and their main features.  The paper leans heavily on the uses for machine learning to protect mobile networks from attacks. Interestingly, this paper separates into two distinct parts with regards to the surveying of papers, coded as the Offensive Perspective, and the Defensive Perspective.  
\newline
The focus on Adversarial Machine Learning, as opposed to Generative Adversarial Networks, for network intrusion detection systems mean that, while this paper is very in-depth and covers a variety of machine learning models including GANs, our paper fits within the gap between AML for IDS and GANs for malware research.
\newline

The survey, Navidan et al., 2021 \cite{Navidan.2021}, covers familiar ground in the research surveyed, but in exceptional depth. The authors are quick to note that, as is evident by the content of these surveys, while GAN for cybersecurity is a relatively new field, it is already being extensively researched. However, this paper has a very small related works section, with only two other surveys mentioned briefly. 
\newline
The survey covers some interesting areas in which GANs are currently being utilised. In the paper \cite{li2019dynamic}, the survey authors note the creation of an ingenious GAN model for morphing traffic flow data, called FlowGAN. The purpose of this model is to train it as to what benign or normal traffic flow patterns look like. Then it morphs the traffic data that needs to pass by undetected, into something with the right features and patterns to be labelled as benign/normal traffic. Such a model would be of interest in regards to malware research as a method to defend against malicious files and applications being disguised as normal and benign traffic. 
\newline
In Future Work, the authors of the survey note that improving ways to avoid image translations are of great significance. This would ease the ability of researchers when finding or developing a new method to use with a different type of data, making it not so cost-expensive with regards to overhead.
\newline

We have found a space within these existing surveys to fill gaps with regards to how Generative Adversarial networks are used in areas of malware research for cybersecurity. We have done so with a focus on creating an overview that will be of use to those individuals in need of a primer on GANs and their potential applications in research into malware. We have attempted to balance depth with breadth of content, and to point readers to other papers that may give them further information on the use of GAN models in these research areas. While our paper might be considered of a parallel topic to \cite{Arora2022433}, the latter paper is much more compact, and as well as dealing more broadly in cybersecurity as a whole, it involves a detailed case study which leaves it little room to discuss the topics in the depth which we have attempted in this survey. Overall, we believe that we have contributed valuable commentary on the state of the art in GAN models for malware research.

\begin{table*}
    \begin{tabular}{|c|c|c|c|}
        \hline
        Topics & Berman et al., 2019\cite{Berman2019} & Liu, et al., 2022\cite{Liu2022123} & Navidan et al., 2021\cite{Navidan.2021} \\
        \hline
        Malware &\cellcolor{gray} X & \cellcolor{gray} X & \cellcolor{gray} X \\
        \hline
        Adversarial Examples & &\cellcolor{gray} X &\cellcolor{gray} X \\
        \hline
        Data Augmentation &\cellcolor{gray} X &\cellcolor{gray} X &\cellcolor{gray} X \\
        \hline
        Network Data &\cellcolor{gray} X &  &\cellcolor{gray} X \\
        \hline
        Reinforcement Learning & &\cellcolor{gray} X & \\
        \hline
        Unseen Examples& &\cellcolor{gray} X & \\
        \hline
        Offensive/Attacker Models &\cellcolor{gray} X &\cellcolor{gray} X &\cellcolor{gray} X \\
        \hline
        Defensive/Defender Models & & \cellcolor{gray} X & \\
        \hline
        Social Network Analysis &\cellcolor{gray} X & &\\
        \hline
        Android Malware & & &\cellcolor{gray} X \\
        \hline
        Financial Fraud Detection  & & &\cellcolor{gray} X \\
        \hline
        Image Enhancement &\cellcolor{gray} X & &\\
        \hline
        Domain Generation Algorithms &\cellcolor{gray} X & &\\
        \hline
        Botnet Detection &\cellcolor{gray} X & & \\
        \hline
        Drive-By Download Attacks &\cellcolor{gray} X & &\\
        \hline
        Password Attacks & & &\cellcolor{gray} X \\
        \hline
        Mobile Network Attacks  & & &\cellcolor{gray} X \\
        \hline
        Internet of Things Attacks & & &\cellcolor{gray} X \\
        \hline
        \hline
        \multicolumn{4}{|c|}{\textbf{GAN MODELS DISCUSSED IN SURVEY}}\\
        \hline
        VanillaGAN & \cellcolor{gray} X &\cellcolor{gray} X &\cellcolor{gray} X \\
        \hline
        CGAN & & & \cellcolor{gray} X\\
        \hline
        DCGAN &\cellcolor{gray} X & &\cellcolor{gray} X  \\
        \hline
        WGAN & & \cellcolor{gray} X & \cellcolor{gray} X \\
        \hline
        BiGAN & & & \cellcolor{gray} X  \\
        \hline
        CycleGAN & & & \cellcolor{gray} X \\
        \hline
        AC-GAN & & & \cellcolor{gray} X \\
        \hline
        MalGAN & &\cellcolor{gray} X & \\
        \hline
        ISGAN & & & \cellcolor{gray} X \\
        \hline
        InfoGAN & & &\cellcolor{gray} X \\ 
        \hline
        FlowGAN & & &\cellcolor{gray} X \\
        \hline
    \end{tabular}
\caption{Types of GAN Research in Related Works}
\label{table:relatedresearch}
\end{table*}

\section{Measuring Performance}\label{metrics}
In most papers related to GAN schemes, there are expected metrics for evaluating a machine learning systems like this one\cite{eberhart1990performance}. The most popular are listed here, as are the ways these show the performance of the GAN.
\subsection{Evaluation Metrics}
The TP, FP, TN, FN scores are often tabulated as a Confusion Matrix to show the performance of the ML algorithm. An example Confusion Matrix can be found in Table
\ref{table:metrics}.
\begin{table*}
\large
\centering
\begin{tabular}{ c c c c c } 
\hline
\multicolumn{4}{c}{}{Predicted Classification}\\
\hline
 & & Benign & Malicious\\
\hline
\multirow{2}{1pt}{}{Actual Classification} & Benign & TP & FP \\   & Malicious & FN & TN\\ 
\hline
\end{tabular}
\caption{Metrics in Machine Learning Classifiers}
\label{table:metrics}
\end{table*}
\subsubsection{True Positive}
The \textit{True Positive/TP} is the number of correctly predicted positive results, or the total number of correctly classified benign samples.

\subsubsection{False Positive}
The \textit{False Positive/FP} is the number of incorrectly predicted positive results, or the total number of incorrectly classified benign samples.

\subsubsection{True Negative}
The \textit{True Negative/TN} is the number of correctly predicted negative results, or the total number of correctly classified malicious samples.

\subsubsection{False Negative}
The \textit{False Negative/FN} is the number of incorrectly predicted negative results, or the total number of incorrectly classified malicious samples.

\subsubsection{Accuracy}
The \textit{accuracy} is the average of correct predictions - of both positive and negative varieties - when classified. Thus, it is the correct predictions divided by the total predictions, or:
\begin{equation}
    acc = \frac{TP + TN}{TP + TN +FP + FN}
\end{equation}
This is the assumed metric in papers or articles which talk only about averages and score. 

\subsubsection{Precision}
Also known as \textit{Positive Predictive Value} or PPV, this is the samples that were classed correctly as benign over all samples that have been classified as benign.
\begin{equation}
    p = \frac{TP}{TP + FP}
\end{equation}

\subsubsection{Recall}
The recall, also known as \textit{true positive ratio, or sensitivity}, is the ratio of samples classed as benign over the total samples classed as benign.
\begin{equation}
    r = \frac{TP}{TP + TN}
\end{equation}

\subsubsection{F1-Score}
This is the \textit{Harmonic Mean} of the \textbf{precision} and the \textbf{recall} values. A \textit{harmonic mean} is one of three types of Pythagorean averages. It is heavily influenced by the lowest of the values, when applied to real numbers, meaning it holds an important place to check the minority classes' accuracy.§
\begin{equation}
    F_1 = 2\left(\frac{pr}{p + r}\right)
\end{equation}

\subsubsection{Inception Score}
When $g$ is the Generator, $d$ is the Discriminator, and there are two finite, label sets, $\Omega_X$ and $\Omega_Y$. As such, $p_g$ is a distribution over $\Omega_X$. $p_Y:\Omega_x\rightarrow M(\Omega_Y)$ is the discriminator function and $M(\Omega_Y)$ is the set of all possible probability distributions over the set $\Omega_Y$. Any image can be $x$, while $y$ is any label. Thus, writing $p_d(y|x)$ is calculating the probability that the image $x$, has the label $y$ - as calculated by the Discriminator Network. The below shows the equation for calculating the Inception Score over all probability distributions, $\Omega_X$, and $\Omega_Y$\cite{Barratt.2018}. 
\begin{equation}\label{inceptionscore}
    IS(g,d) = \exp (\mathbb{E}_x\sim p_g[D_{KL}(p_{d}(\cdot \mid x) \parallel \int p_d(\cdot \mid x)p_g(x)dx)])
\end{equation}
There is a pre-trained network which measures the Inception Score, and the higher the score of the model, the higher the quality of the images produced \cite{salehi2020generative}. The Inception Score and Network were introduced in 2016 for Convolutional Neural Networks by Szegedy et al. \cite{szegedy2016rethinking}. It was originally developed to remove human subjectivity in computer vision research.

\subsubsection{Mode Score}
The Mode Score is meant to be an improved version of the Inception Score. It still measures the quality and diversity of images, but it counts the prior distribution of labels\cite{Borji.2018}.
\begin{equation}\label{modescore}
    MS(\mathbb{KL)} = \exp (\mathbb{E}_x[\mathbb{KL}(p(y \mid x) \parallel p(y^{train})]-\mathbb{KL}(p(y) \parallel p(y^{train})))
\end{equation}

\subsubsection{Fréchet Inception Distance}\label{frechet}
There is another equation derived from the Inception Score. The Fréchet Inception Distance (FID) and the Inception Score (IS) together can be used as an attempt to solve overfitting\footnote{Overfitting occurs in statistical analysis when too few samples are present and the model is fitted too closely to this small selection of samples, making its ability to generalise low.}. The FID is shown below in Equation \ref{fideq}. The purpose of the FID is to examine the distance between groups. It was also developed for the specific task of image processing in machine learning \cite{szegedy2016rethinking}. Frechet Inception Distance for any two probability distributions, $\mu$ and $\nu$, over the set of real numbers, $\mathbb{R}^n$, is calculated as follows:
 \begin{equation}\label{fideq}
     d_F(\mu,\nu) := \left( \underset{\gamma\in\Gamma(\mu,\nu)}{inf}\int_{\mathbb{R}^n\times \mathbb{R}^n}\parallel x-y \parallel^2d\gamma(x,y) \right)^{1/2}
 \end{equation}
 The set used in the FID here, $\Gamma(\mu,\nu)$ is actually the 2-Wasserstein distance over $\mathbb{R}^n$\cite{Borji.2018}.
There is a second calculation for the FID score, but it works only over two Gaussian, multi-dimensional distributions, $\mathcal{N}(\mu,\sum)$ - symbolised below as $r$ - and $\mathcal{N}(\mu',\sum')$ - shown below as $g$.
\begin{equation}
    FID(r,g) =  \parallel \mu_r - \mu_g \parallel _2^2 + Tr(\sum_r + \sum_g - 2(\sum_r\sum_g)^{1/2}) 
\end{equation}

\section{Dataset}\label{dataset}
The different datasets on which the Machine Learning algorithms are trained have a significant effect on how they read the given data, what their biases or preconceived ideas may be, and how they are trained to recognise different integral features. In this section we have attempted to cover the main datasets used in the papers we have surveyed. 

\subsection{DGArchive}\label{dgarchive}
The DGArchive is a set of domains, of 43 families, classes, or variants, with more than 20 million domains as of 2015\cite{ploh15}. These domains are from models in Domain Generating Algorithms which create domains for Control \& Command centres for botnets. The database of malicious botnet C\&C domains allows for machine learning classifiers to be trained on how to detect domain name malware.
This data is extremely important in creating new machine learning methods for identifying botnet C\&C centres (as in \cite{Choudhary.2019}). The compilation of this information into such a large and comprehensive database is an important research tool.
The DGArchive dataset is also used to create adversarial machine learning models, such as MaldomDetector\cite{Almashhadani.2020}, which undertake the generation of malicious domain names itself, and allows researchers to test defensive machine learning algorithms on an adversary. 

\subsection{VirusTotal}\label{virustotal}
This repository of both anti-virus software and a database of files, both benign and malicious, is known as VirusTotal. It can scan a given file using 70 antivirus systems as well as checking with URL and domain blacklisting programs\cite{virustotal22}. Each uploaded file - as well as resulting in a report stating the findings and results labelling it either benign or malicious and how these results were arrived at - is also kept and added to the overall database of VirusTotal files. 
The service is free for research and non-commercial use, and licences can be purchased for commercial users or those needing a large sample set of data\cite{Anderson.2018}. In addition to scanning, users can request a subset of the database for use in training and testing their own algorithms. This is a often used service in machine learning research, such as \cite{Zhao.2018}, because of the depth and breadth of malware covered by the VirusTotal dataset. It is also useful because of the constant updating the servers get as users upload their own programs and files to scan. This is an unusual dataset in that regard, where other datasets covered in this section are static and set, while VirusTotal is continually changing and updating.
VirusTotal contains files, programs, Android applications\cite{Canbek.2018}, applications for Windows, Mac, Linux, iOS, and so on. This is another point in favour of the database - the type of data available for training and testing is extensive and covers a lot of ground, where other datasets discussed only cover one type of information.

\subsection{Contagio}\label{contagio}
Contagio is a publicly available dataset of malware, specifically samples of Android malware and benign applications\cite{Singh.2021y8e}. The dataset was updated periodically between 2011 and 2018, and can be found for open access online at a \href{http://contagiodump.blogspot.com}{range of places}\cite{milacontagio}.
The fact that this dataset is focused on Android malware makes it extremely useful, as overall, openly available databases of mobile malware are not as prevalent as those for desktop malware or traffic flow data. As of 2021, Contagio contained 11,960 malicious and 16,800 benign samples of Android software\cite{Singh.2021y8e}, with a total size of approximately 9GB\cite{337}. This dataset can be accessed in .zip format for researchers and white-hat activities.

\subsection{Drebin}\label{drebin}
Drebin is a repository for Android malware, similar to Contagio. Drebin contains 123,453 applications and 5,560 malicious APKs for Android, in a variety of malware families, totalling about 6GB in size\cite{337}. It was collected from 2010 to 2012. It was originally proposed as part of a paper in which Drebin - a new algorithm - was proposed to catch malware on Android smartphones. As part of this, a database of 5,560 malicious Android APKs were collected, which now make up the Drebin dataset \cite{Arp.2014}. It is important, therefore, to differentiate between the Drebin static-analysis detection software, and the Drebin dataset. Both were organised around the use of eight main feature sets for analysis \cite{Arp.2014}. These sets are as follows:
\begin{itemize}
    \item Hardware components
    \item Requested permissions
    \item App components
    \item Filtered intents
    \item Restricted API calls
    \item Used permissions
    \item Suspicious API calls
    \item Network addresses
\end{itemize}

\subsection{Comodo}\label{comodo dataset}
The Comodo Database\cite{comodo2022}, maintained by Comodo Antivirus, contains sample files of malware. As part of a program to encourage research, Comodo partnered with universities and launched Comodemia\cite{comodemia2022}. This gave access to Comodo's tools to researchers internationally, for research purposes only.
The Comodo malware database primarily contains files classified as unknown malware - totalling 147,103 instances. The other categories are Trojan viruses (462 instances) and Unwanted Applications (13 instances). It is a clearly unbalanced dataset. However, like the other datasets, it is used to train and test different machine learning classifiers, as in \cite{Ye.2018}.

\subsection{VirusShare}\label{virusshare}
The VirusShare database\cite{9vs} is a large online, open-source repository for malware. A user account is required for access, but anyone with an account can access and download the live viruses in the database. The database is found at VirusShare.com, and is maintained by Corvus Forensics, though anyone can submit files to be added to the dataset.
Some researchers, like \cite{Duzgun.2021}, in which 14,616 unique examples were taken from the VirusShare database, have taken portions of the VirusShare database and melded them with other datasets in order to balance and augment datasets as necessary. In \cite{Menendez.2019}, the VirusShare dataset was augmented with malware obtained from Kaggle, in order to test the author's proposed malware detection scheme, and was used as a benchmark when the new model was run against VirusTotal's (see Section \ref{virustotal}) antivirus detection program.

\subsection{Microsoft Malware Classification Challenge (2015)}\label{microsoftviruschallenge}
The Microsoft Malware Classification Challenge \cite{https://doi.org/10.48550/arxiv.1802.10135}, made available an open source database of malware examples for Windows. The challenge was part of a general push to coders into creating their own deep learning methods for malware classification \cite{Chivukula.2021}. It can be found primarily on Kaggle\footnote{Dataset and competition information can be found here: \href{https://www.kaggle.com/c/malware-classification}{https://www.kaggle.com/c/malware-classification}}. It was an open competition on the site, for teams to come together and develop their own solutions to the challenge. It was completed by 377 teams during the open competition time between April 13 - 18, 2015. 
The dataset contains more than 20,000 malware examples, and according to the authors of the challenge \cite{Ronen.2018}, as of 2018 it had been cited in over 50 research papers. It is now a widely used dataset for machine learning research, as in \cite{7856826}, \cite{8268747}, \cite{Kalash.2018}.

\subsection{MNIST}\label{mnist}
The MNIST dataset, or Modified NIST dataset, a collection of images of handwritten characters, was introduced in 1998 by LeCun et al\cite{Lecun.1998}, for the primary purpose of computer vision and recognition tasks in machine learning. It contains characters which are clear representations, as well as those which have been perturbed to examine the extent to which a computer vision algorithm can recognise deformed figures.
Many models dealing with the challenges of computer vision in machine learning have utilised this dataset. For example, the InfoGAN model, discussed in Section \ref{infogan}, was trained and tested on the MNIST dataset, \cite{Chen.2016}, before moving on to 3D renderings. Another study, surveying the effectiveness of different models of GANs, used the MNIST dataset to benchmark the performance of each model \cite{Cheng.2020}. 
Since its inception, new, updated versions of MNIST have been proposed. One such dataset is EMNIST (Extended MNIST), which takes the dataset from digits only into handwritten alphanumeric characters \cite{Cohen.2017}, with a total of 814,255 samples in all classes combined.

\section{Types of GAN models}\label{GANs}

The papers we have surveyed have used a range of variants of the traditional GAN. As a reference and refresher, we have included this section. in which we cover the different types of GANs we will be discussing, and the points of difference in each. We also wanted to clearly illustrate the issues inherent in the standard GAN model, so that the variations which are developed specifically for overcoming them are understood. 

\subsection{Vanilla GAN}\label{vanilla gan}
The traditional, or Vanilla, Generative Adversarial Network is the original proposed model from Goodfellow et al's 2014 paper \cite{goodfellow2014generative}. The authors proposed the GAN model as an alternative to Variational Autoencoders for adversarial machine learning.
This original version of the Generative Adversarial Network is a deep learning model, and was based on adversarial nets as a framework, with back-propagation. This model uses a two-player minmax game to adjust the weights, as per Figure  \cite{goodfellow2014generative}. An important distinction is that while the discriminator has access to both real and generated data, the generator has \textit{no access to either}, and so has to rely on the value functions and the back-propagation to change the weights and take the model closer to producing realistic generated output \cite{Cheng.2020}. The generator and discriminator are both able to be non-linear mapping functions \cite{Mirza.2014}. The GAN model came through the adversarial nets framework, a way of dealing with weights without Markov chains, and instead using back-propagation\cite{Mirza.2014}.
For an in depth overview of how the Goodfellow GAN operates, please refer to Section \ref{howganworks}.\newline
\textbf{Inherent Problems in the Goodfellow GAN Architecture}
\subsubsection{Mode Collapse Problem}\label{mode collapse} The complexities of the MinMax game that are essential to the standard/Vanilla GAN result in an optimisation problem. This is solved in the standard version using the gradient descent-ascent (GDA) method \cite{Durall.2020}. However, this can lead to serious errors in convergence resulting in failure of the GAN, including a problem known as \textit{mode collapse}\cite{Thanh-Tung.2020}. Combating this problem is one of the reasons there are so many variations of GAN models - many are developed to help avoid the convergence problems in optimisation of the minmax function as much as possible.

\subsubsection{Catastrophic Forgetting}\label{catastrophic forgetting} Catastrophic Forgetting (CF) occurs when \say{knowledge of previously learned tasks is abruptly destroyed by the learning of the current task} (\cite{Thanh-Tung.2020}). CF can prevent proper convergence in the model, and limit it from finding the necessary local maxima optimum for the task it is set. 
Remembering the location and features of the real samples used to train the generator is essential - when the generator loses these samples, catastrophic forgetting occurs as the new generated samples are not created with the real samples as a guide\cite{Thanh-Tung.2020}.

It is important to note that mode collapse and catastrophic forgetting are interlinked - they make the other worse in situations where both problems arise. The equation for describing the optimal
Discriminator in Goodfellow's GAN model is shown in Equation \ref{optimal discriminator}. The training criteria for a given discriminator, $D$, and a generator, $G$, are shown in Equation \ref{trainingdg}.

\begin{equation}\label{optimal discriminator}
    D^*_G(x)=\frac{p_{data}(x)}{p_{data}(x)+p_g(x)}
\end{equation}
\newline
\begin{align}\label{trainingdg}
    V(G,D) & = \int_xp_{data}(x)log(D(x))dx\\
    & +\int_xp_z(z)log(1-D(g(z)))dz\\
    V(G,D) & =\int_xp_{data}(x)log(D(x))dx\\
    & +p_g(x)log(1-D(x))dx
\end{align}

\subsection{Conditional GAN}\label{cgan}
The Conditional GAN (CGAN), proposed in \cite{Mirza.2014}, modify the original vanilla GAN. In the original model, the generative process could not be controlled or conditioned. It was unsupervised entirely. CGANs allow the generation process to be controlled and directed, meaning that the model can be steered towards a focus on a particular class, or feature. The original paper proposing a CGAN model utilised the MNIST dataset (see Section \ref{mnist}) to test its capabilities.
The change in control occurs when the focus is put on some element $y$, which can be a class, value, feature, so on, and this $y$ is fed into both the generator and discriminator as an additional layer of input. In the original proposal, the generator is fed not only the focus $y$, but also a noise function, $p_z(z)$ \cite{Mirza.2014}.
In a subsequent study, a CGAN for facial recognition was proposed \cite{Gauthier.2014}, which used sampled random noise for $p_z(z)$ and a random sampling for $y$ which is taken from the training dataset, utilising a Pazan window, $p_y(y)$.

\subsection{Deep Convolutional GAN}\label{dcgan}
The Deep Convolutional Generative Adversarial Network, or DCGAN, was proposed in a 2015 paper titled "Unsupervised representation learning with deep convolutional generative adversarial networks" \cite{radford2015unsupervised}. The model for DCGAN was based in research around convolutional neural networks, and how they might offer opportunities for growth in other machine learning models. The paper was focused on the generation of sudo-natural images, as GAN models are so highly effective in image generation tasks. While most deep learning algorithms are black-box methods, it is possible through careful tuning to examine the underlying functions of a Convolutional Neural Network (CNN) model. The authors made use of several changes to traditional CNN architecture, from the following papers:
\begin{itemize}
    \item \textit{Striving for simplicity: The all convolutional net}\cite{springenberg2014striving}
    \item \textit{Inceptionism: Going deeper into neural networks} \cite{mordvintsev2015inceptionism}
    \item \textit{Batch normalization: Accelerating deep network training by reducing internal covariate shift}\cite{ioffe2015batch}
\end{itemize}

\subsection{cDCGAN}\label{cdcgan}
The conditional Deep Convolutional Generational Adversarial Network, or cDCGAN, takes the properties of both the CGAN (See section \ref{cgan}) and DCGAN (see section \ref{dcgan}) models. It was proposed as part of a paper focusing on handwritten Arabic characters \cite{Mustapha.2022}, a significantly more complicated task than identifying English alphanumeric characters. Arabic characters are distinctly different in that Arabic lettering can have similar characters to the extent that they "are only distinguishable by dots"\cite{Mustapha.2022}. The database utilised was the AHDBase/MADBase \cite{Abdleazeem.2008}, containing 70,000 digits, chosen because it was the database that best matched the MNIST database of numeric handwritten digits. (\ref{mnist}).
The discriminator in the cDCGAN model is a deep CNN. The Leaky activation function is used in this model. The generator matches its picture quality to 32x32 pixels, with a LeakyReLu function.

\subsection{Bi-directional GAN}\label{bigan}
The Bi-directional Generative Adversarial Network (BiGAN) was proposed in a 2017 paper called "Adversarial feature learning"\cite{donahue2016adversarial}. Similarly to MGAN (see Section \ref{mgan}) this is a three party model, consisting of an encoder, a generator, and a discriminator. The role of the encoder is to map data $x$ to a latent space representation $y$. Donahue et al specify that the encoder is taught to invert the generator, even though the modules do not interact with one another or directly process the other module's outputs. \newline
The BiGAN model is meant to excel at tasks that involve semantic data and representation. They are also an entirely unsupervised model in machine learning.
Interestingly, BiGAN was brought into the spotlight in Bioinformatics in a 2021 paper titled "BiGAN: LncRNA‐disease association prediction based on bidirectional generative adversarial network"\cite{Yang.2021}, the BiGAN model proved highly effective. When compared against the three gold-standard algorithms for detecting the "associations of IncRNA-disease pairs"\cite{Yang.2021}, BiGAN achieved the highest scores, including 93.1\% for the AUC. That was several percentage points higher than the standard methods. BiGAN models are now found in many different fields, including research into malware.

\subsection{MalFox}\label{malfox}
The creation of MalFox, a GAN model for creating attacks and new malware \cite{Zhong.2020} gave an important and powerful tool to those testing or attacking existing systems. MalFox is an amalgamation of parser, generator, and discriminator layers, which takes as input Windows Portable Executable files, or PEs, and outputs the same executables. This makes it a more practical tool that the more common adversarial example generators which often take an image created by a feature extraction process and don’t create functioning malware in pre-approved file types. 
Since its inception, MalFox has undergone more than one transformation, but even the original version shows the power of GAN-based schemes for attack purposes. The initial experiments were used against pre-trained classifiers – Decision Tree; Random Forest; Logistic Regression; Support Vector Machine; Multi-Layer Perceptron; Vote; Long Short-Term Memory; Bi-directional LSTM; LSTM Average; Bi-LSTM Average; LSTM Attention; and BiLSTM Attention. The evasion rate – the percentage of times the program was classified as benign by the systems - MalFox achieved was 99\% minimum across the board. This is a stunning display of the power of these schemes. Furthermore, when tested against the open-access giant VirusShare, the detection rate was only 29.7\% on average. The evasion rate on the same was averaged as 56.2\%. MalFox and the experiments done show exactly how powerful these schemes can be.

\subsection{MalGAN}\label{malgan}
In 2017, the authors \cite{Hu.2017} created a GAN which proposed black-box adversarial examples for attacking via Windows binaries. This scheme, called MalGAN, is now widespread, with multiple different variants and branches of development. The use of binaries, and portable executable files, is one of the ways this is so successful at showing the potential of GAN-based attacks. The authors of the original MalGAN were able to get the detection rate down to almost zero. This clearly demonstrates the danger that is posed by GAN attack systems. Since its inception, MalGAN has been modified and improved, under the auspices of creating the best database for the training of robust detection schemes. \cite{Wang2021Zhang} proposed using an LSGAN model to address weaknesses like mode collapse. The model they propose still uses the MalGAN scheme, after the use of the LSGAN method, which involves implementing a Least Square function. Their purpose was making MalGAN more robust and avoiding the potential fallout from limiter problems or mode collapse. The authors are still focused on creating adversarial examples, though they focus more on poisoning attacks as well as the traditional ‘perturbation’ attacks, which change only a small portion of the code while still retaining the capabilities or functionality of the original sample. They are not the only researchers to build a version of MalGAN with Least Square functions to increase the robustness of the function. \cite{Wang2021Mi} also suggested the use of a Least Square function in order to minimize mode collapse. The authors of this paper also changed up the activation functions, and added LeakyReLU to the mix. They achieved an 85\% success rate over seven different ML classifiers.

\subsection{Least Square GAN}\label{lsgan}
Mao et al propose the creation of a GAN variant called the Least Square GAN, (LSGAN)\cite{Mao.2017}. Named for its innovation, the model uses the Least Squares equation for the discriminator. This helps to minimise the Pearson divergence of $X^2$. The Pearson $X^2$ Divergence is a variant of the $f$-divergence,  The LS function can help distance correctly classified samples from the genuine data, improving the performance of the classifier, and thus increasing the training level of the generator. The objective functions of the LSGAN model are presented in Equation \ref{lsgan objective functions}, The objectives functions for the LSGAN model, from \cite{Mao.2017}.

\begin{align}\label{lsgan objective functions}
     \underset{D}{min}V(D)_{LSGAN} & = \frac{1}{2}\mathbb{E}_{x\sim p_{data}}[(D(x)-b)^2] + \frac{1}{2}\mathbb{E}_{x\sim p_z(Z)}[D(G(z))-c)^2] \\
     \underset{G}{min}V(G)_{LSGAN} & = \frac{1}{2}\mathbb{E}_{x\sim p_{z}(z)}[(D(G(z))-c)^2]
\end{align}

\subsection{AC-GAN}\label{ac=gan}
The auxiliary classifier GAN (ACGAN) was proposed in 2016 by Odena et al \cite{Odena.2016}. The ACGAN variant was proposed, at the time, for use in image generation, but has since moved into other subjects, as many GAN models do due to their easily transferable nature.
The variant distinguishes itself by the use of an auxiliary decoder network \textit{within} the discriminator. As a result, the algorithm can:
\begin{itemize}
    \item Give as output the label of the class for training samples.
    \item Output a subset of the set of latent variables used to generate the samples.
\end{itemize}
According to Navidan et al. \cite{Navidan.2021}, the strongest point of difference between the ACGAN model and the CGAN variant is that in order to determine class labels, the CGAN relies on the conditioning of the generator. In contrast, the ACGAN predicts class labels due to the auxiliary decoder network. The way the ACGAN predicts class labels can be found in Equation \ref{acgan class predictions}.
\begin{align}\label{acgan class predictions}
    L_S & = E_{x\sim p_{data}(x)}[\log D(x)] + E_{z\sim p_z(z)}[\log (1-D(G(z)))]\\
    L_C & = E_{x\sim p_{data}(x)}[\log Q(c|x)]+E_{z\sim p_z(z)}[\log Q(Q(c|z))]
\end{align}

\subsection{IS GAN}\label{is-gan}
The Identity-Sensitive Generative Adversarial Network was proposed and focused on \textit{face photo-sketch synthesis} \cite{Yan.2021}. This is the process by which a photo of a face is turned into a sketch through machine learning. The goal of creating the ISGAN model was to create a formal image translation task that addresses the problem of turning a photo into a pseudo-hand drawn sketch. This area does have a security and police angle - on occasion, when given a poor quality image of the face of a suspect, turning it into a sketch can help to idetify features that may not be as prominent in the original photo. Especially when machine learning is involved, allowing the ISGAN to understand and augment the original image. 
ISGAN is not the only GAN model that has been applied to this task, but, when the benchmark tests were run against the current state-of-the-art methods, ISGAN was either on par or above them in score\cite{Yan.2021}.

\subsection{InfoGAN}\label{infogan}
The Information Maximising Generative Adversarial Network, or InfoGAN, model was first proposed in Chen et al, in 2016 \cite{Chen.2016}. The authors noted the ability of InfoGAN's model to untangle images of handwritten characters. The model was tested and trained using the MNIST dataset (see Section \ref{mnist}). It was also utilised on 3-dimensional images of faces, and on pictures of house street numbers.  
In performance, the InfoGAN model adds 'negligible' complexity to the vanilla GAN (\ref{vanilla gan}) model. The training itself was based on the training done for a DCGAN (\ref{dcgan}), rather than a vanilla GAN.

\subsection{fvGAN}\label{fvGAN}
In \cite{Li202038}, the authors develop a method by which they can utilise GAN to build malicious code into PDF files in such a way as to evade detection by even schemes dedicated to the detection of PDF malware. The proposed method, feature vector GAN or fvGAN, took in the fact that features in PDF files are highly interconnected and interdependent, meaning one cannot simply change the features to the ones required. First they had to pull out the features that were most essential. Using \textit{mimicus}, an invention of their own design, they were able to pull feature vectors with 135-dimensions from the files. Once the fvGAN had been trained on the Contagio and Surrogate datasets (from the original PDFRate study\cite{laskov2014practical}) of both malicious and benign PDFs, they used it to create PDFs with content injection attacks to great effect. 

\subsection{CycleGAN}\label{cyclegan}
The CycleGAN model is foremost an image translation mechanism \cite{Zhu.2019}. Cycling an 'unpaired' image from one domain to another is its main purpose. Proposed in 2017 by Zhu et al \cite{Zhu.2017}, CycleGANs have become a reliable tool in image processing, and have assisted researchers in many domains. With regards to security, it is important to note that CycleGAN models can be used for biometrics - particularly facial recognition. More recently, a CycleGAN variant was proposed for video-to-video translations - Mocycle-GAN \cite{Amsaleg.2019}. This raises the possibility of using this type of GAN to build facial recognition into CCTV software.

\subsection{ProGAN}\label{progan}
The Proximity Generative Adversarial Network, or ProGAN \cite{gao2019progan}, is meant to preserve the proximities of instances and samples that are reduced in dimensionality. The original proximity in the space prior to dimensionality reduction must be preserved, and thus ProGan was created.
The proximity of nodes in this subject is classed as first-order, second-order, and so on. If a node is connected to another node with an edge, it is considered first-order. These relationships are preserved through network embedding.

\subsection{MGAN}\label{mgan}
Mixture Generative Adversarial Networks (MGAN) was proposed focusing on overcoming the mode collapse problem in vanilla GAN models (see Section \ref{mode collapse}). This problem is a serious risk for standard GAN models. MGAN seeks to address that issue by using multiple generators to create generated output based on the real data given to the discriminator \cite{hoang2018mgan}. The generators are trained simultaneously, rather than sequentially, and the resulting distributions can be \textit{mixed} to achieve a realistic distribution. The ultimate goal is to create a three-party minmax game, rather than the traditional two party game with vanilla GAN. The parties involved in a MGAN minmax game are: the discriminator, the classifier, and the set of generators. The different generators are meant to work harmoniously, and the authors point out the importance of having the different generators specialise in different data modes.

\begin{landscape}\label{GANTable}
\centering
\begin{table}
\resizebox*{1.6\textheight}{\textwidth}
{\begin{tabular}{|c|c|c|c|c|c|c|c|c|c|c|c|c|}
\hline
GAN Models & Balancing datasets & Attack \& Security & Malware & Functional Malware & Adversarial Examples & Malicious Traffic & Feature Extraction & Phishing URLs & Behaviour Tracking \\
\hline
Goodfellow GAN\cite{goodfellow2014generative}   &\cellcolor{gray} X & \cellcolor{gray} X & \cellcolor{gray} X &\cellcolor{gray} X & \cellcolor{gray} X & & & & \\
\hline
CoGAN & & & \cellcolor{gray} X & & \cellcolor{gray} X & & & & \\
\hline
DCGAN & \cellcolor{gray} X & \cellcolor{gray} X &\cellcolor{gray} X & \cellcolor{gray} X & \cellcolor{gray} X & & & & \\
\hline
ALI-GAN & & &\cellcolor{gray} X & & \cellcolor{gray} X & & & & \\
\hline
CoRGAN & & &\cellcolor{gray} X & & \cellcolor{gray} X & & & & \\
\hline
CoRAGAN & & &\cellcolor{gray} X & & \cellcolor{gray} X & & & & \\
\hline
Sequential GAN &\cellcolor{gray} X & \cellcolor{gray} X & \cellcolor{gray} X & \cellcolor{gray} X & \cellcolor{gray} X & & & & \\
\hline
GenAtSeqGAN & & & & & & & & & \cellcolor{gray} X \\
\hline
LSTM GAN & &\cellcolor{gray} X & \cellcolor{gray} X & \cellcolor{gray} X & \cellcolor{gray} X & & & & \cellcolor{gray} X \\
\hline
Bi-objective GAN & & \cellcolor{gray} X & \cellcolor{gray} X & \cellcolor{gray} X & \cellcolor{gray} X & & & & \\
\hline
WGAN-GP & & & \cellcolor{gray} X & \cellcolor{gray} X & \cellcolor{gray} X & & & \cellcolor{gray} X & \\
\hline
MalGAN & & \cellcolor{gray} X & \cellcolor{gray} X &\cellcolor{gray} X & \cellcolor{gray} X & & & & \\
\hline
CGAN & & & & & & & & & \\
\hline
fvGAN & & \cellcolor{gray} X & & & & &\cellcolor{gray} X & & \\
\hline
TrafficGAN & & \cellcolor{gray} X & \cellcolor{gray} X & & & \cellcolor{gray} X & & & \\
\hline
Mal-IGAN & \cellcolor{gray} X & & \cellcolor{gray} X & \cellcolor{gray} X & \cellcolor{gray} X & & & & \\
\hline
n-gram MalGAN & & & & & & & & & \\
\hline
IDSGAN & \cellcolor{gray} X & & & &\cellcolor{gray} X & \cellcolor{gray} X & & & \\
\hline
iKnight & & & & & \cellcolor{gray} X &\cellcolor{gray} X &\cellcolor{gray} X & & \\
\hline
PDF-GAN & & &\cellcolor{gray} X &\cellcolor{gray} X &\cellcolor{gray} X &\cellcolor{gray} X & & & \\
\hline
GAN-tCDGAN &\cellcolor{gray} X & &\cellcolor{gray} X & &\cellcolor{gray} X & & & & \\
\hline
IDS-GAN & & & & & &\cellcolor{gray} X & & & \\
\hline
Mal-LSGAN & &\cellcolor{gray} X &\cellcolor{gray} X &\cellcolor{gray} X &\cellcolor{gray} X &\cellcolor{gray} X &\cellcolor{gray} X & & \\
\hline

\end{tabular}}
\label{fig:usesofgan}
\caption{The different types of GAN models surveyed in this paper with their most common uses.}
\end{table}
\end{landscape}

\section{Areas of Use}\label{areasofuse}
There are many areas in which Generative Adversarial Networks are of use, and these include a selection of cybersecurity related topics. A broad overview of the types of use each different model of GAN is used in is shown in the table in Section \ref{GANTable}.
\subsection{Classification and Images}\label{classificationimages}
One of the first, and enduring, tasks for which GANs are used is that of image classification\cite{zhu2018generative}. The ease with which GANs can compare and create images with the necessary similarities and contextual elements, meaning the creation of many images that belong to clear classes, is a task that GAN models do so well even non-image based tasks are often translated into images\cite{Kargaard2018411} for the ease it provides when creating augmented datasets using GAN schemes. The implications for the us in malware are self-evident. The ability to generate new samples of malware families in order to train machine learning based detection schemes on what the \textit{features} of a family of malware are, is a leap forwards in terms of defensive technology. for an example, in \cite{Moti.2019}, the authors use deep learning GAN models to generate unseen malware examples and train schemes against the signatures of these new malware images. While the authors did not achieve a high increase in classification, the principle of their work has been examined by others as well. One study, \cite{Lu.2019} found they could increase classification accuracy of malware samples by 6\% through the generation of synthetic examples of malware for training purposes.
In \cite{f2}, authors implemented GAN for the classification of greyscale images that were created by transforming malware files with feature extraction. This task is suited to GAN schemes because GAN, more so than any other problem space, excels at image classification. In this study, the classifier's performance improved by approximately 6\%. In \cite{Wang20203775}, the authors again translate malware files into greyscale images in order to use GAN models on them for classification. GAN systems are excellent at picking out images that have significant similarity, which in this case means that they belong to the same malware families. Using the Microsoft Malware Classification Challenge \cite{82n}, the system is run against AE-SVM \cite{yousefi2017autoencoder}, tDCGAN \cite{kim2018zero}, Strand \cite{drew2016polymorphic}, and MCSC-asm \cite{ni2018malware}, It performs better when classifying the malware family the images belong to than these state of the art classifiers, with the lowest error rate.

\subsection{Data Augmentation, Rare Classes, and Balancing a Dataset}\label{whyaugmentation}
Machine learning models suffer from a desperate need for training data. The amount of data needed to train complex systems to the point at which they know how to deal with the data coming into their systems is enormous. And, unless they are an unsupervised learning model, that data requires logging and labelling. This task is an enormous undertaking, and one that requires human input, hours upon hours of sitting at computers and labelling each piece of data that will be sent to the model for training and verification processes. GAN is an effective tool to potentially solve some of the issues which arise out of a need for data augmentation\cite{Mimura2020}.
Instead of an individual creating hashes of new malware as it arrives, GANs can be used to generate and generalise synthetic malware examples for machine learning models. Rather than learning through the hashed values of malware files, the GAN can produce images of malware files and rare families for training purposes. The importance of feature extraction for learning in malware detection is of particular weight in this scenario. This is shown clearly in \cite{Lu.2019}, in which they augment their malware datasets with synthetic samples created by GAN models.
There are other problems with the data necessary for machine learning models too. 
\newline
Augmenting an existing dataset such that the different types of data, the different classes, are able to be trained for the highest levels of accuracy is highly important in classifying data. It is a major use of GAN schemes (see \cite{mariani2018bagan,Tan.2020,Zhang2022900,Wang2021,10.32604/cmc.2022.029858,Guo2021,undefined, Alghazzawi20223877}).  Taking a dataset that is perhaps too small, or has classes that are too imbalanced to learn robustly, is an area GAN models shine. As an example, \cite{Moti.2020} focuses on using GAN models to create new samples of different classes of malware in order to balance a dataset on which to train machine learning models.
\newline
In many existing datasets, especially those related to cybersecurity, there exists a significant imbalance in the classes of data within a dataset. There are many sets where the class ratios are significantly imbalanced.  As seen in \cite{Chen2021}, a dataset with very unbalanced classes can be made more even across types using a GAN to solve the rarity of certain classes. The involved datasets were malware designed as Android APKs, which the authors translated into greyscale images. By supplementing these datasets of Android malware with GAN methods, the authors were able to achieve increases of 5-20\% in the F1-score. This shows the power of GAN models in augmenting rare data. \newline

\subsection{Zero Day Malware}\label{adversarialexamples}
The accurate and speedy classification of malware files and malicious code in computer systems is a expansive task, and one that has remained an enduring problem in the cybersecurity domain\cite{kim2018zero}. This is a task that is getting bigger and more pressing, not losing significance. The ease with which malicious code can be edited once antivirus software has been updated to detect that particular type of malware means that the creators of this malware have the edge in the battle between attackers and defenders. The sophisticated attacks using polymorphic malware (see \ref{polymophismevolution}, which morphs and changes itself in order to escape detection, make the task of identifying the malicious code even more difficult\cite{drew2016polymorphic}. Traditional antivirus software relies on the use of hash codes. Once the authors of the antivirus software have identified a piece of malicious code, it is hashed into a value string that specifically identifies that piece of code. The antivirus definitions are updated, and the program knows how to recognize that code, should it come into contact with the computer on which the defense software is installed. This does, however, rely on the idea that this malware has been identified before it turns up on the target system. Because the slightest change in the malicious code causes cascading changes in the hash code, all a malware developer needs to do to escape the antivirus system is move a portion of code to a different place within the program. This retains functionality, while changing the hash code of the program enough that it will no longer be picked up by traditional antivirus schemes. Until the antivirus developers find and identify the new variant of the malware, the traditional scheme will not notice the altered malware and it will be able to operate undetected. GAN has become a tool for building new adversarial examples that the defender can train with to prevent this outcome\cite{Kang20214105}. Similarly, in \cite{10.1109/access.2021.3056482}, the authors address this problem using GAN-based adversarial examples to train their blockchain method of intrusion detection. Their LSTM-CGAN model for generating these examples allowed the resulting classifiers to jump several percentage points in accuracy when the classifier was trained on the AEs as opposed to when it was trained on the unenhanced dataset. \newline
One of the bigger stumbling blocks in malware detection is the unseen or ‘zero-day’ attacks\cite{bilge2012before}. Because of the types of data that a GAN model can create, generating new adversarial examples and zero day malware is a possible way to train a security model on unseen data. In \cite{Liu2020}, the authors use a GAN model they name TrafficGAN to create new malicious traffic patterns for zero-day attacks, partially by including noise into the data as substitution for some unseen traffic, In \cite{Moti2019319}, a standard variation GAN with a LeakyReLU activation function is implemented to train IDS models on unseen malware. While their contribution on the whole is an increase of only 1\% accuracy overall, the idea of using the data generated by GAN to increase the robust nature of a model on data it hasn't seen in training sets is a useful and practical one. In fact, the authors of \cite{Hu.2017} use GAN modelling to create black-box attack methods. Their chosen model in particular, MalGAN (see Section \ref{malgan}), was created entirely towards the view that GAN schemes could create functional and unseen examples of malware to attack machine learning systems. The attacks they conducted on non-neural network systems managed to achieve a True Positive Rate of zero, while on the neural network based models (RF, LR, DT, SVM, MLP, VOTE), which had achieved a TPR of 92\% minimum prior to the attack tests, managed to achieve a TPR of 0.19\% maximum when the generated attack data was integrated into the testing set. This study shows how vulnerable IDS or malware detection systems are to the data and attacks generated by a malicious GAN. On the other had, when the authors of \cite{Umer.2021} generated their own zero-day attacks and added them into the training datasets for their IDS, they were able to achieve a success rate of 84\% when classifying unseen malware. Utilising the Microsoft Malware Classification Challenge (2015) dataset \cite{82n}, they outperformed 14 other state-of-the-art systems, and achieved a 98\% success rate when classifying the zero day attacks. This shows that using GAN to create unseen examples and zero-day attacks can be a powerful force for creating truly robust classifiers and detectors. 

\subsection{Detection Evasion}\label{detectionevasion}
Given the success at generating unseen examples in order to train more robust systems, the use of GAN methods to create malicious code which evades detection is a logical step. Building GAN schemes which can generate malicious files that evade current detection schemes is necessary for the training of the next generation of malware detection schemes.
In \cite{Zhu.2021o7q}, the authors create a scheme using GAN to evade detection from a broad range of machine learning models - Multi-layer Perceptron, Decision Tree, Logistical Regression, Support Vector Machine, Random Forest - and found that feature extraction was the key in making their scheme achieve the minimal True Positive Rate (TPR). As they increased the selection of features in both attacker and defender, the TPR rose. While the authors take the steps of adding benign features into the malicious examples they create, they achieve a slightly less impressive TPR than the previously mentioned scheme - the lowest TPR is under 11\%. Unlike the previous model, the authors utilise n-gram feature selection, a method borrowed primarily from Natural Language Schemes, but which removes the step of translating the malware into a different form - such as the popular method of making malware 'images' - and allows the model to work on the raw data itself.
The creation of the MalFox GAN scheme in \cite{Zhong.2020} shows the ability of GAN schemes to create black-box attacks, in which the attacker knows nothing about the actual structure of the system they are attacking. In a black box attack, the attack must be generalised enough that it can be employed successfully against any defender. In MalFox, which was trained specifically with Obfusmal, Stealmal, and Hollowmal as techniques to perturb the data for attacking, when used against the online malware repository Virus Share \cite{9vs}, the detection rate was minimised to 56.2\%.   
In \cite{Li.2021}, the authors use a standard GAN model against a Deep Neural Network defence system in order to create Windows malware that is perturbed just enough to evade detection and retain its original purpose and function. The authors achieved this using raw byte sequences and training the GAN on existing byte sequences of malware. They achieved evasion rates of more than 50\%. 

\subsection{Applied Attack GAN Models}\label{attackmodels}
The flipside of using GAN to generate unseen malware and use it to train antivirus and IDS models, is the use of GAN to create highly effective, theoretical attacks on existing systems. The possibilities offered by GANs for creating new ‘adversarial examples’ to use against detection systems, mean that there is significant opportunity to build new malware with the same functionality, but in a way that is fast and avoids detection with high rates of success. 
As we discussed in Section \ref{GANs}, the creation of models like MalGAN and MalFox (see Section \ref{malgan} and \ref{malfox}), are clear demonstrations of the attack potential of GAN models in generating malware. In \cite{Li.2018p6b}, the authors managed to build a model which evaded firewalls in order to attack Android systems with a success rate of 95\%. This is a troubling note for security against new malware creation techniques. The use of these systems to \textit{train} the new generation of malware detection schemes should be therefore a high priority for security researchers and developers.

\section{Discussion}\label{discussion}
As popular as GAN research is, there are almost as many types of proposed models for GAN as there are papers discussing GANs. There are some which can be easily seen and categorised as types of the same genus, and then there are those rare new examples, which offer a fresh method for implementing GANs within a research or work setting. Surveying as many papers proposing solutions to difficulties in particular sectors results in surveying almost as many problems in others. We have gathered some particular points of interest and problems which may offer future lines of research or simply act to temper future models with an eye towards creation of new malware detection schemes.

\subsection{Malware and the Sandbox}\label{malwaresandbox}
One issue with identifying malware based on features and contextual relationships is that it may require the malware be run before it can be accurately identified. This type of analysis is \textit{dynamic} analysis, whereas detection systems which use the file without running it are performing \textit{static} analysis. The majority of the papers we have assessed in this study propose static analysis \cite{Zolkipli.2011}. However, this then causes concern for the implementation of these methods - can they run successfully in real-time, stopping the malware from running even before it has been classified? This puts a level of vulnerability into the scheme. What damage will be done while the machine identifies the currently executing program as malware? A mitigation for this type of problem is to run the programs in a \textit{sandbox} before classification. This means they cannot affect the performance of the system or exploit any vulnerabilities prior to identification\cite{lindorfer2011detecting}. Another suggestion is that the program is not run until the binaries have been extracted and used to try to identify the malware family to which it belongs. This problem is not specific to Deep Learning or Machine Learning based detectors - it has been of concern for many years, and has often been addressed using the sandbox option. However, this simply lead to the creation of "environmental detection" in malware\cite{yokoyama2016sandprint}, allowing the code to detect when it was being run in a sandbox and when it was in the real system environment.  

\subsection{Polymorphism, Evolution and the Dangers of Malware}\label{polymophismevolution}
Malware developers have taken stock of the current state of research in the area and used it to their advantage. New types of polymorphic malware, which twists and turns itself into code-based pretzels in order to avoid detection, are able to fool signature-based detection systems \cite{Zolkipli.2011}. GANs schemes have the ability to perturb the code of existing malware code and use these unseen examples to train new models. This ability has potential to defend against polymorphic malware. 
One method, discussed in \cite{Zolkipli.2011}, is to use malware \textit{behaviour} to teach new detection systems. Of course, this has innate risk, because the malware has to be executed in order to complete the behaviour analysis. In such a case, the idea would be to run the malware sample in a 'sandbox', as discussed in Section \ref{malwaresandbox}. Another potential problem is fighting the \textit{method by which the malware is spread.} Social engineering hacking, sending malicious files to email, or using watering hole attacks\footnote{A watering hole attack involves infecting a site the target is likely to visit, rather than directly attacking the target's device. See \cite{nq7} for a detailed examination of waterholing attacks.} to download the malware onto the target computer \cite{Gazet.2010} are all tasks that are reliant on the ways in which the user interacts with their devices. Ideally, a new and improved machine learning based detection scheme would protect users even when they clicked on email links or went to download a PDF from a new site. This makes training effective and accurate classifiers are priority in cybersecurity, and one of the preeminent areas of malware research. 

\subsection{Optimisation and Nature Inpired Computing}\label{optimise}
In all machine learning models based on Neural Networks, there is a need to optimise the model. The number of layers, the nodes in each layer, the activation function, the weights. All of these make the model significantly better or worse at its task, and for the most part, they are altered and improved through trial and error. One group of researchers, however, has taken this optimisation problem and made it the focus of their research into GAN models \cite{Du2020}. Using Genetic Algorithms, they run through the different options for optimisation automatically, with the non-denominated sorting genetic algorithm (NSGA-II) finding the best values for the GAN model. This true positive score of the optimised GAN was more than 98\% on the MNIST dataset, introduced in \cite{lecun1998gradient}, which is made up of images of handwritten letters. The same format for optimisation was employed when the model ran on the malware dataset taken from the Vision Institute \cite{nataraj2011malware}, and achieved a true positive rate of 97.87\%, a score several points higher than the GAN run on the same dataset without the optimisation of the NSGA-II algorithm. This suggests there is a road to take for optimising the layout and structure of each of the GAN models, and that it is possible another Nature-Inspired Computing area may be how that is achieved. As such, it is an interesting direction for malware researchers to explore.

\subsection{Data Format and Translation}\label{data format}
In many of the examples profiled in this paper, the data, be it traffic flow, machine language, API calls, or malware files, is often translated into different types of images. This makes the job of the GAN model easier because they work so well with image classification tasks, but it also means that there is additional complexity in the algorithms due to the necessity of translating the data into the correct format. There are papers, however, such as \cite{Moti.2020}, \cite{Sur2020}, \cite{Hu.2017}, \cite{Zhong.2020}, \cite{Yumlembam.2022}, or \cite{Li202038} which use the raw data, without translating it to a type more palatable to a GAN scheme. This is important because of the overhead of these different models. When investigating to find an efficient model for use in a given domain, ensuring that there is little to no additional or unnecessary overhead seems a significant consideration.
\newline
One thing common amongst almost all papers discussed is the need to change the form in which the data is used. The pre-processing, referred to in \cite{Alavizadeh.2022} as \textit{data triage}, is a cost-expensive and high overhead requirement in both time and processing power. There are those that have managed to avoid translating the data overmuch - such as the research presented in \cite{Liu2020} - but most translate data into images or carefully sectioned bytes of data. This is an area of concern while real-time application of a GAN based IDS or firewall is a goal. To further the idea of a real-time GAN scheme, the methods of data pre-processsing need to be carefully examined. They offer unacceptably large overheads for real world application. 
Because malware files are generally simple and popular to translate into images, the task is not as arduous as it might be in other areas of interest. 

\subsection{Ethics and Responsibility in Malware Research}\label{ethics}
The ethical questions posed in \cite{mirjalili2018semi} offer an interesting path for future research - there is a general lack of discussion in GAN research papers about the ramifications of possible misappropriation or misclassification of the work contained within. This is an important step to consider in all academic research, and the ethical implications of things like Deepfake\cite{westerlund2019emergence}, or AdvandMal\cite{Wang2021Mi} are likely to cause extreme and often unintended effects. Whether by GAN or VAE, the type of research in papers which create new models for malware synthesis are an excellent example of why researchers need to be careful in balancing the quest for knowledge and the security/ethical risks of their research. Knowing this type of attack is possible, and with the results they achieved, is important for those who develop defensive mechanisms for this type of attack, but it is also useful to black hat individuals and malicious operators. The amount of detail over the model created, how it operates, and was trained is where that balance of ethics and information comes into play.
These things must be looked at carefully and applied with consideration. The situation of neural network applications as they are demands that we take stock of what has been built, how the dataset has been labelled, how the features have been established, and how/where the neural network is going to be deployed. While this work is starting to take place in the field - see \cite{matthias2004responsibility},\cite{sand2022responsibility} - there is much still to cover, and researchers like those in \cite{martin2019designing} are offering frameworks for incorporating ethics into the fabric of machine learning research.
Therefore, like many research topics, it has a flipside, and both sides need serious ethical review and consideration to ensure that the research benefits those who are most vulnerable.

\section{Future Research Directions}\label{futureresearch}
There are still many avenues for potential research. The methods employed by the authors in \cite{Zhu.2021o7q} to avoid the popular step of translating the dataset into sequences or images and instead working on the data directly using the n-gram feature extraction method is certainly an area worthy of future research for more applications. The incredibly low detection rate achieved by 
The use of Genetic Algorithms to optimise the performance metrics of GAN models \cite{Du2020} is an avenue which could prove very fruitful. The development of real-time, dynamic analysis and detection is a challenge researchers are still only beginning to scratch the surface of \cite{Zolkipli.2011}, and requires further research into the types of secure environments in which this analysis can safely take place.
The realm of malware research contains so many possible avenues for research when it comes to GAN algorithms, and this has been illustrated in this paper to the best of our ability. 

\section{Conclusion}\label{conclusion}
This paper has presented a wide range of research in the current malware research space, using different types of Generative Adversarial Models. Our aim is to have provided an explanation of not only what Generative Adversarial Networks are and how they are trained and assessed, that we have also given an effective grounding in the applications within the malware research community, which GANs may work with both in the current literature and in any potential future research. To that end, we have iterated through the explanation of GAN's basic functions; the work on other surveys done in related areas, particularly to demonstrate that there has yet to be an in-depth survey paper on the uses of GAN in cubersecurity and malware; we have explained the different metrics used for evaluation; we gave an in-depth review of the datasets currently favoured in the problem space; we included a list on the different GAN models that are discussed throughout the survey; we have delved in depth to the areas of use that are currently most popular, we have endeavoured to provide a discussion and survey that goes into detail so as to give the reader the full picture; and we have presented potential future avenues for research in the area. We hope that this survey of malware research through the lens of Generative Adversarial Networks, and the way in which they can be employed, has given the reader an idea of where to start with their own research in this area, or given them an update state of the art. The field of GANs for malware research is only getting started, and there is much to do, and many questions to be answered.


\begin{backmatter}
\section*{Declarations}
\subsection*{Competing interests}
  The authors declare that they have no competing interests.

\subsection*{Funding}
  The authors would like to thank the Ministry of Business, Innovation, and Employment (MBIE) from the New Zealand Government to support our work with the grant (MAUX1912) which made it possible for us to conduct the research.

\bibliographystyle{bmc-mathphys} 
\bibliography{bibliography}      




\section*{}
\begin{table}
Appendix A: List of Abbreviations \\
\resizebox{!}{.39\paperheight}{
    \begin{tabular}{|l|l|}
         \hline
            \cellcolor{lightgray}\textit{Abbreviation} & \cellcolor{lightgray}\textbf{Description} \\
            \hline
            AC-GAN & Auxiliary Classifier GAN \\
            \hline
            ACC & Accuracy \\
            \hline
            AE & Adversarial Example \\
            \hline
            AE-SVM & Auto-encoder Support Vector Machine \\
            \hline
            ALI-GAN & Adversarially Learned Inference GAN \\
            \hline
            AML & Adversarial Machine Learning \\
            \hline
            API & Application Programming Interface \\
            \hline
            APK & Android application format \\
            \hline
            AUC & Area Under the Curve \\
            \hline
            Bi-LSTM & Bi-dierctional Long Short-Term Memory \\
            \hline
            BiGAN & Bi-directional Generative Adversarial Network \\
            \hline
            C\&C / CC & Control and Command Server \\
            \hline
            cDCGAN & Conditional Deep Convolutional GAN \\
            \hline
            CF & Catastrophic Forgetting  \\
            \hline
            CGAN & Conditional Generative Adversarial Network \\
            \hline
            CNN & Convolutional Neural Network \\
            \hline
            CoRGAN & Correlation-Capturing GAN \\
            \hline
            DCGAN & Deep Convolutional Generative Adversarial Netwrk \\
            \hline
            DGA & Domain Generation Algorithm \\
            \hline
            DL & Deep Learning \\
            \hline
            DNN & Deep Neural Network \\
            \hline
            DNS & Domain Name Sever \\
            \hline
            DT & Decision Tree \\
            \hline
            EMNIST & Extended MNIST database (see MNIST) \\
            \hline
            FID & Fréchet Inception Distance \\
            \hline
            FN & False Negative \\
            \hline
            FP & False Positive \\
            \hline
            fvGAN & Feature Vector Generative Adversarial Network \\
            \hline
            GA & Genetic Algorithm \\
            \hline
            GAN & Generative Adversarial Networks \\
            \hline
            GDA & Gradient Descent/Ascent \\
            \hline
            GenAtSeqGAN & Generalized Attentive Sequential Generative Adversarial Network \\
            \hline
            IDS & Intrusion Detection System \\
            \hline
            IDS-GAN / IDSGAN & Intrusion Detection System GAN \\
            \hline
            InfoGAN & Information Maximising Generative Adversarial Network \\
            \hline
            IS & Inception Scare \\
            \hline
            ISGAN & Identity-Sensitive Generative Adversarial Network \\
            \hline
            LeakyReLu & Leaky Rectified Linear Unit \\
            \hline
            LR & Logistical Regression \\
            \hline
            LSGAN & Least Squares Generative Adversarial Network \\
            \hline
            LSTM & Long Shhort-Term Memory \\
            \hline
            LSTM GAN & Long Short-Term Memory GAN \\
            \hline
            LSTM-CGAN & Long Short-Term Memory Conditional Generative Adversarial Network \\
            \hline
            Mal-IGAN / MaliGAN & Maximum-Likelihood Augmented Discrete Generative Adversarial Net- works \\
            \hline
            Mal-ISGAN & (See ISGAN, MalGAN) \\
            \hline
            MCSC-asm & Machine Classification using SimHash and CNN \\
            \hline
            MGAN & Mixture Generative Adversarial Network \\
            \hline
            ML & Machine-Learning \\
            \hline
            MLP & Multi-Layer Perceptron \\
            \hline
            MNIST & Modified National Institute of Standards and Technology Database \\
            \hline
            MS & Mode Score \\
            \hline
            NIC & Nature-Inspired Computing \\
            \hline
            NSGA-II & Non-Denominated Sorting Genetic Algorithm  \\
            \hline
            PPV & Positive Predictive Value / Also known as Precision \\
            \hline
            ProGAN & Proximity Generative Adversarial Network \\
            \hline
            RF & Random Forest \\
            \hline
            SGAN & Semi-Supervised Generative Adversarial Network \\
            \hline
            SVM & Support Vector Machine \\
            \hline
            tCDGAN & Transferred Deep-Convolutional Generative Adversarial Network \\
            \hline
            TN & True Negative \\
            \hline
            TP & True Positive \\
            \hline
            TPR & True Positive Rate \\
            \hline
            VAE & Variational Autoencoders \\
            \hline
            WGAN & Wasserstein Generative Adversarial Network \\
            \hline
            WGAN-GP & Wasserstein Generative Adversarial Network with Gradient Penalty \\
            \hline
    \end{tabular}}
    \label{tab:abbreviations}
\end{table}

\end{backmatter}
\end{document}